\documentclass[conference]{IEEEtran}

\usepackage{graphicx}
\usepackage{booktabs}
\usepackage{array}
\usepackage{amsmath}
\usepackage{amssymb}
\usepackage{caption}
\ifCLASSINFOpdf
\else
\fi
\begin{document}
%
\title{A Novel Speech-Driven Lip-Sync Model with CNN and LSTM}


\author{\IEEEauthorblockN{Xiaohong Li, Xiang Wang, Kai Wang and Shiguo Lian}
\IEEEauthorblockA{AI Innovation and Application Center, \\China Unicom,
Beijing, China}}


%


\maketitle

\begin{abstract}
Generating synchronized and natural lip movement with speech is one of the most important tasks in creating realistic virtual characters. In this paper, we present a combined deep neural network of one-dimensional convolutions and LSTM to generate vertex displacement of a 3D template face model from variable-length speech input. The motion of the lower part of the face, which is represented by the vertex movement of 3D lip shapes, is consistent with the input speech. In order to enhance the robustness of the network to different sound signals, we adapt a trained speech recognition model to extract speech feature, and a velocity loss term is adopted to reduce the jitter of generated facial animation. We recorded a series of videos of a Chinese adult speaking Mandarin and created a new speech-animation dataset to compensate the lack of such public data. Qualitative and quantitative evaluations indicate that our model is able to generate smooth and natural lip movements synchronized with speech.
\end{abstract}

\begin{keywords}
speech-driven lip-sync; facial animation synthesis; convolutional neural network; LSTM
\end{keywords}

%
\IEEEpeerreviewmaketitle

\section{Introduction}
Creating lifelike and emotional digital human has wide applications in many fields, such as film and games, facial repair and therapy, child education and so on. Nowadays, it is often seen a digital human serving as a newscaster or a narrator. Facial movement, especially the lip movement when speaking, is one of the most important component for digital human to express themselves. The lip shape must match the pronunciation, otherwise it will make audience feel uncomfortable or even fake, leading to the uncanny valley effect. In recent years, a lot of research on speech-driven lip-sync has been carried out. These works can be divided into two categories: 2D and 3D, depending on whether the output is 2D video or a 3D animation. In this work, we focus on the 3D facial animation generation, which is widely used in video games and films.

Although various works on 3D facial animation have been proposed in recent years~\cite{Shimba 2015,Li 2017,Cudeiro 2019}, few works have addressed the issue of animating a 3D model speaking Chinese due to the lack of open dataset and adapted algorithm. To compensate the lack of training data, we used a web camera to record several hours of the video of a male Chinese character speaking Mandarin, with rate of 60 frames per second (fps). The recorded data includes images and the synchronized voice. Afterwards, we created a 3D mesh model of a human head, which was used to transfer the facial motion captured from the human performer to a digital face with the help of a facial tracking software and a retargeting software from Faceware Technologies~\cite{Faceware}.

To effectively make use of the collected data, we designed a deep neural network which takes speech feature as input and outputs the 3D vertex displacement. The speech feature is extracted with a pre-trained speech recognition model, and the vertex displacement is used to drive 3D mesh models to generate accurate facial movements synchronized with the input speech sound. The proposed network combines a one-dimensional convolution and LSTM (Long Short-Term Memory) and is able to generate realistic, smooth and natural facial animations. Note that although we only collected the voice and motion of a male character as training data, the facial animation generated by our model can be applied to 3D virtual characters of different styles and genders. In addition, our model is robust to the speech of different people with different voices.

The remaining part of the paper is organized as follows: Section~\ref{sec:review} reviews the related work of speech-driven facial animation methods. The details of the data collection and the proposed network is introduced in Section~\ref{sec:method}. Section~\ref{sec:res} shows the experimental results and the conclusion is given in Section~\ref{sec:con}.

\section{Related Work}\label{sec:review}

Speech-driven facial animation has received intensive attentions in the past decades. Existing methods can be classified into viseme-driven and data-driven approaches. The viseme-driven approaches adopt a two-phase strategies: speech recognition algorithms are first used to segment speech into phonemes, which are then mapped to visual units a.k.a visemes. Contrarily, the data-driven approaches learnt from a large amount of data and learnt a model which directly transfers from speech or text to facial animations.

Mattheyses et al. declared that an accurate mapping from phonemes to visemes should be many-to-many due to the coarticulation effect~\cite{Mattheyses 2013}. They introduced a many-to-many phoneme-to-viseme mapping scheme using tree-based and k-means clustering approaches, and achieved better visual result. Edwards et al. depicted the many-valued phoneme-to-viseme mapping using two visually distinct anatomical actions of jaw and lip, and proposed the JALI viseme model~\cite{Edwards 2016}. They first computed a sequence of phonemes from the speech transcript and audio, and then extracted the jaw and lip motions for individual phonemes as viseme action units and blended the corresponding visemes into coarticulated action units to produce animation curves. At last, a viseme compatible rig was driven by the computed viseme values. Zhou et al. applied the JALI model and further introduced a three-stage LSTM network to predict compact animator-centric viseme curves\cite{Zhou 2018}.

As it is difficult to accurately identify phonemes and the coarticulation effect is difficult to simulate with simple functions, more and more researchers preferred the data-driven methods. Shimba et al. trained a regression model from speech to talking head with LSTM network and applied lower level audio features instead of phonemes as input~\cite{Shimba 2015}. The output of their network was Active Appearance Model (AAM) parameters \cite{Cootes 2001}. Suwajanakorn et al. also used a recurrent neural network to learn the mapping from raw audio features to mouth shapes based on many hours of high quality videos of the target person~ \cite{Suwajanakorn 2017}. They synthesized mouth textures and composed them with proper images to generate a 2D video that matched the input speech. Unlike the above 2D methods, Cudeiro et al. proposed an audio-driven 3D facial animation method~\cite{Cudeiro 2019}. They trained a neural network on their 4D scans dataset captured from 12 speakers, and used a subject label to control the generated speaking style. In addition, they integrated the speech feature extraction approach DeepSpeech~\cite{Hannun 2014} into their model to improve the robustness to different audio sources. However, their trained model is only applicable to targets of the FLAME model~\cite{Li 2017}, which is a statistical head model. Meanwhile, their model performs better on English than other languages because their training data only includes English speech.

\section{Method}\label{sec:method}

When dealing with temporal and sequential tasks, such as speech recognition, machine translation and text processing with relevance to the context, the Recurrent Neural Networks (RNNs) are often used considering its advantage over the traditional feed-forward neural networks which cannot exhibit temporal dynamic behavior. The RNNs are a class of neural networks that allow previous output be used as input to the recurrent layer. However, it is difficult to solve problems that require learning long-term temporal dependencies using standard RNNs due to the vanishing gradient phenomenon. Therefore, a memory cell that is able to maintain information for long period is introduced, which is known as the Long Short-Term Memory (LSTM) unit. In this work, we use LSTM as the backbone network to map the speech feature extracted with DeepSpeech to vertex offsets. Specifically, we use unidirectional LSTM to model this mapping and make further improvements on this basis.

A common LSTM unit contains a cell, an input gate, an output gate and a forget gate. The cell remembers information from previous intervals and the three gates control the memorizing process. Let $x_t$ denote the input of a LSTM unit at time $t$. The activation of this unit is updated as follows:

\begin{gather}\label{eq:lstm}
f_t=\sigma(W_f\cdot[h_{t-1},x_t]+b_f) \\
i_t=\sigma(W_i\cdot[h_{t-1},x_t]+b_i) \\
o_t=\sigma(W_o\cdot[h_{t-1},x_t]+b_o) \\
\tilde{C}_t=tanh(W_C\cdot[h_{t-1},x_t]+b_C) \\
C_t=f_t*C_{t-1}+i_t*\tilde{C}_t \\
h_t=o_t*tanh(C_t),
\end{gather}

where $f$, $i$, $o$, $C$, $h$ represent the forget gate, the input gate, the output gate, the cell state and the cell output respectively. $\sigma$ denotes the sigmoid activation function. $W$ and $b$ are the weights and biases respectively.

Next, we will introduce our dataset generation method, the process of speech feature extraction, our network framework and the loss function in details.

\subsection{Dataset Generation}

To the best of our knowledge, there has not been any public Chinese dataset that can be used for the training of speech-driven 3D facial animation. So we produced our own dataset. We applied the THCHS-30 corpus~\cite{Wang 2015},  which is an open Chinese speech database published by Center for Speech and Language Technology (CSLT) at Tsinghua University, as our speech script. It contains 1000 sentences and covers most of the phonemes in Chinese. A male character was invited to speak these sentences with normal speed in front of a camera. This camera contains a microphone and is able to record synchronized video and sound. It was used to capture the image data at a rate of 60 frames per second and record the surrounding sound signals at the same time. We collected totally four hours of video data.

\begin{figure*}
\centering
\includegraphics[width=0.85\textwidth]{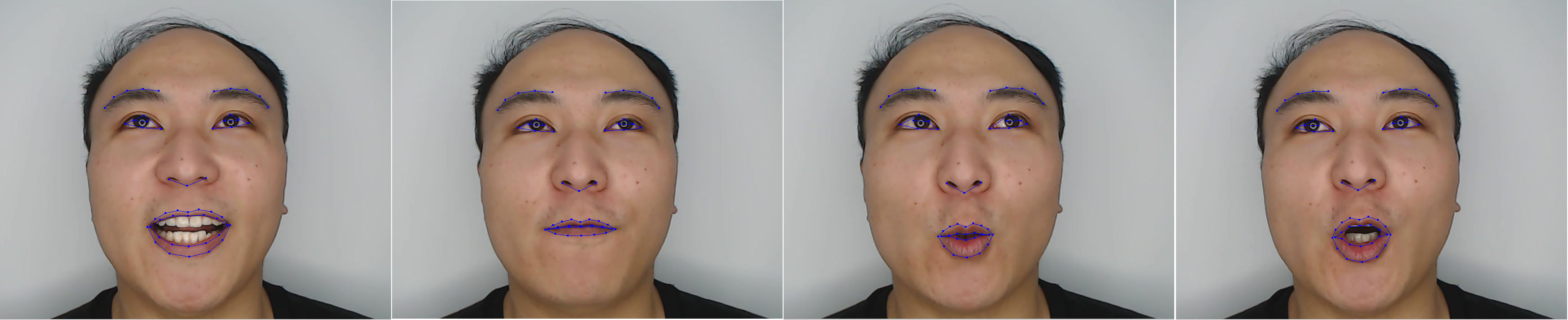}
\caption{Facial landmarks tracked with Faceware Analyzer.}
\label{track-figure}       
\end{figure*}

\begin{figure*}
\centering
\includegraphics[width=0.85\textwidth]{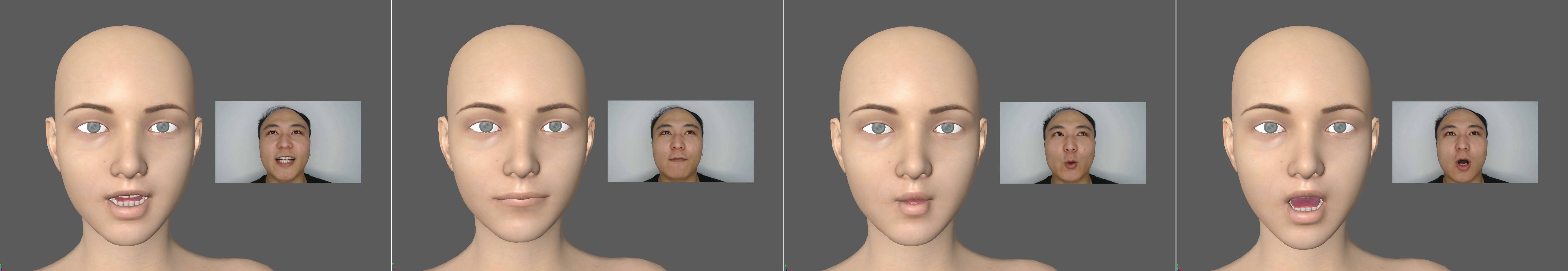}
\caption{Facial motions transferred from video to 3D character with Faceware Retargeter.}
\label{retarget-figure}       
\end{figure*}

\begin{figure}
\centering
\includegraphics[width=0.35\textwidth]{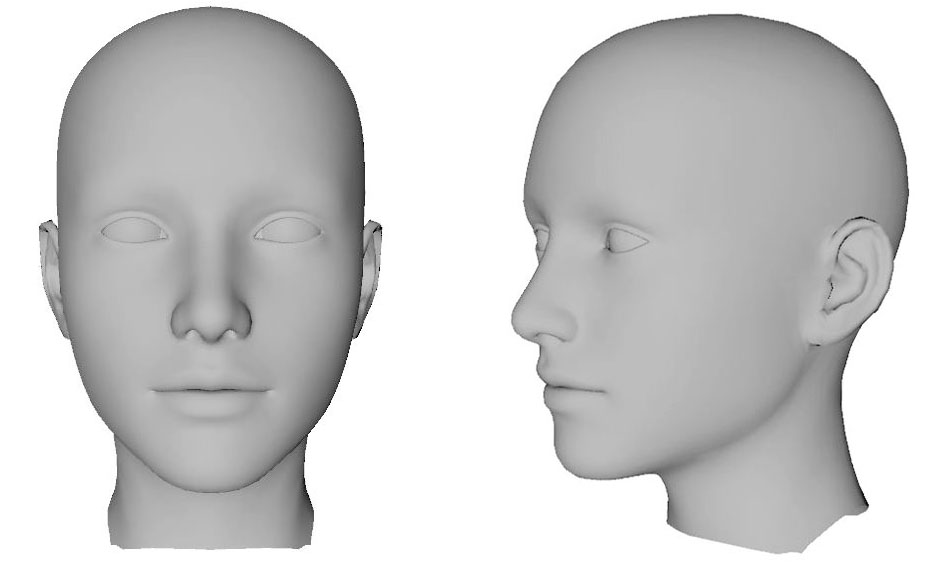}
\caption{3D virtual character used to generate 3D facial animation data.}
\label{template-figure}       
\end{figure}

Next, we utilized two tools from the Faceware software kits~\cite{Faceware} to convert the collected videos into 3D animations. One is the face tracking tool Analyzer, which tracks facial landmarks from video in a markerless way (see Figure~\ref{track-figure}). The other is Retargeter, which is a high quality facial animation solving software that retargets facial motion from a tracked video onto a 3D character. It needs a prepared 3D model with proper rigs and generates facial animations by controlling the rig movement. We used these two tools to transfer the facial animations from the captured video to a 3D character, and obtained the 3D face data by recording the vertex position of the head of the animated 3D character in every frame. Figure~\ref{retarget-figure} shows the retargeting result from video frames to a virtual 3D character. The prepared virtual 3D character is regarded as a template, which is a mesh model in "zero pose" (see Figure~\ref{template-figure}). All the animated 3D faces have the same topology as the template.

We split the generated 3D facial animation data and accompanied audio data into a training set (900 sentences), a validation set (50 sentences) and a test set (50 sentences). The ground truth of the vertex displacement is calculated by subtracting the template from the generated 3D faces.

\subsection{Speech Feature Extraction}

In order to improve the robustness of our model to input speech signals, we use a pre-trained speech recognition model to extract speech feature. Like VOCA~\cite{Cudeiro 2019}, we adapt the DeepSpeech model~\cite{Hannun 2014}, which is trained on hundreds of hours of voice data and generalized to different speakers, voice speed, environment noise, etc. We used the pre-trained model provided at the DeepSpeech GitHub releases page, which used Mel Frequency Cepstral Coefficients (MFCCs)~\cite{Davis 1980} to extract the input audio features, and replaced the fourth recurrent layer with a LSTM unit. The output of the model is a sequence of character probabilities, which is used as the input of our network.

Given an audio clip with a length of $T$ seconds, we first resample the audio to fixed 16kHZ, and calculate its MFCCs. After normalization, we feed them to the pre-trained DeepSpeech model to extract audio features and resample them to 60 fps, which is consistent with the frame rate of videos in our dataset. Finally, the output is a two-dimensional array with size of $60T \times D$, where $D$ is the number of letters in alphabet plus a blank label.

\subsection{Network Architecture}

\begin{figure*}
\centering
\includegraphics[width=0.8\textwidth]{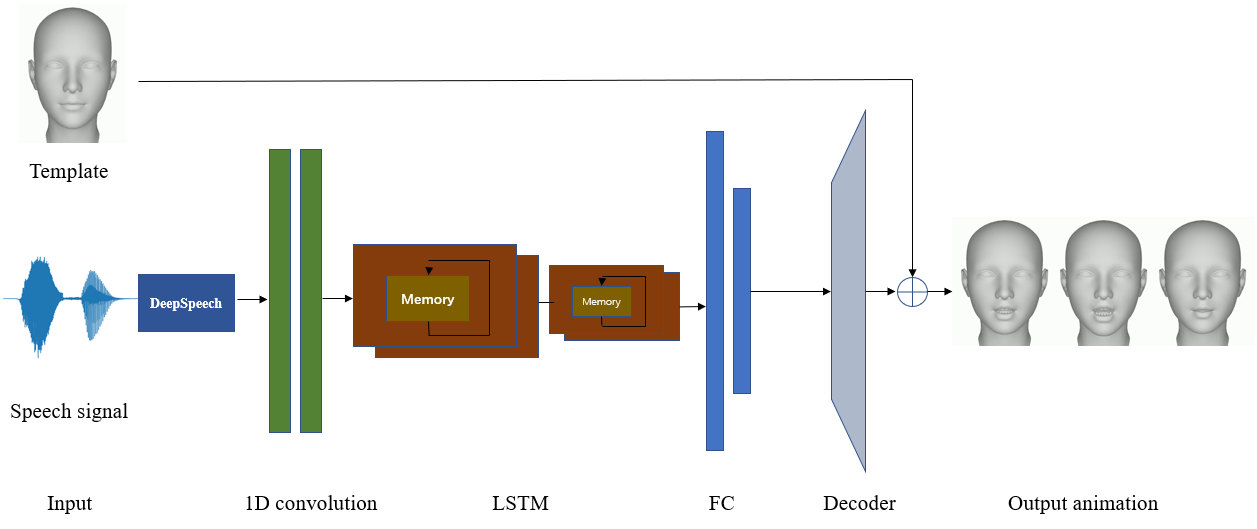}
\caption{Network architecture}
\label{network-figure}       
\end{figure*}

The architecture of our network is inspired by WaveNet~\cite{Oord 2016} proposed by Aaron van den Oord et al., which is a deep generation model of raw audio waveforms. It is able to generate speech that mimics many different speakers, and make it sound more naturally than lots of existing text-to-speech systems. It realized a time sequence generation with only a series of convolutional layers stacked, which is faster to train than the recurrent network. However, its receptive field is limited. To overcome this limitation, we combine the convolutional layers with LSTM blocks to construct our network structure, as shown in Figure~\ref{network-figure}.

The network is implemented using Tensorflow framework~\cite{Abadi 2016}. The first part of the network is composed of two one-dimensional convolutional layers, four unidirectional LSTM blocks and two fully connected layers, which are used together to transform speech features extracted by DeepSpeech to a low-dimensional embedding. The latter part of the network is a decoder consisting of a fully connected layer with linear activation. The decoder maps the embedding into a high-dimensional space of 3D vertex displacement, whose dimension is $5713\times 3$ in our case, as the 3D facial model we used in training set has $5713$ vertices, and the position of each vertex is represented by the Cartesian coordinate in 3D space. Table~\ref{network-table} shows the specific parameters of our network.

\begin{table}
\centering
\caption{The proposed network architecture}
{\begin{tabular}{cccccc} \toprule
\textbf{Type} & \textbf{Kernel} & \textbf{Stride} & \textbf{Output} & \textbf{Activation} & \textbf{Num} \\ \midrule
DeepSpeech & - & - & 29 & - & -  \\ \midrule
Convolution & 5$\times$1 & 1$\times$1 & 32 & ReLU & 2 \\
LSTM & - & - & 128 & - & 2 \\
LSTM & - & - & 64 & - & 2 \\ \midrule
Fully connected & - & - & 128 & tanh & - \\
Fully connected & - & - & 50 & linear & - \\ \midrule
Fully connected & - & - & 5713$\times$3 & linear & - \\ \bottomrule
\end{tabular}}
\label{network-table}
\end{table}

During inference, we take variable-length audio clips as input and output vertex displacement at 60 fps. Then the computed displacement is added to the position of each vertex of a 3D face template to drive the face deformation. The topology of the face template here needs to be the same as that of the one we used for training data generation.

\subsection{Loss Function}

Suppose the vertex displacement sequence output by the network during training is $\{\tilde{y}_t\}$, $t=1..T$, and the ground truth of the vertex displacement is $\{y_t\}$, $t=1..T$, where $T$ is the number of video frames. We train the model by minimizing the following loss function:

\begin{equation}\label{eq:loss}
Loss = \omega_1*L_p+\omega_2*L_v,
\end{equation}
where $L_p$ is the reconstruction loss, $L_v$ is the velocity loss, and $\omega_1$ and $\omega_2$ are the weight coefficients respectively.
$L_p$ is defined in Eq.~\ref{eq:ploss}, in which $F$ represents the Frobenius Norm. It calculates Euclidean distance of vertices between the predicted output and the real facial animation, and thus used to constrain the gap between the predicted vertex coordinates and the ground truth.

\begin{equation}\label{eq:ploss}
L_p = {\parallel y_t-\tilde{y}_t \parallel}_F^2.
\end{equation}

The velocity loss $L_v$ is defined as Eq.~\ref{eq:vloss}. It uses backward finite differences of the mesh vertices of adjacent frames to estimate the deformation speed of the face vertices, and calculates the difference between the predicted value and the ground truth. It has a smoothing effect. When only reconstruction loss is used, the mouth movement is obvious, while lip jitter cannot be avoided. Through qualitative and quantitative analysis, it is found that the velocity loss can reduce the lip shaking and improve the model accuracy significantly.
\begin{equation}\label{eq:vloss}
L_v = {\parallel (y_t-y_{t-1})-(\tilde{y}_t-\tilde{y}_{t-1}) \parallel}_F^2.
\end{equation}

The weight coefficients for $L_p$ and $L_v$ have to be set carefully, otherwise the mouth shape will be inaccurate. We set the weights of the reconstruction loss and velocity loss to $1.0$ and $0.5$ respectively. More details can be found in the follow-up comparative experiments in Section~\ref{sec:res:quan}.

\section{Experiments}\label{sec:res}

We compared our network with other methods and conduct ablation experiments for velocity loss term. Both qualitative and quantitative evaluations and analysis will be introduced in this section.

\subsection{Qualitative Evaluation}

Although we only used the voice of a male for training, our method generalizes well for female and even synthesized voices. The test results for different speech sources can be found in the supplementary video\footnote{https://www.dropbox.com/s/71oayo97aywd3l9/A\%20Novel\%20Speech-Driven\%20Lip-Sync\%20Model\%20with\%20CNN\%20and\%20LSTM.mp4?dl=0}. We compared our network with LSTM and VOCA, whose results are also shown in the supplementary video. We used the pre-trained VOCA model which is claimed to work for any language, and use the same test audio spoken in Mandarin as input.

\begin{figure}
\centering
\includegraphics[width=0.5\textwidth]{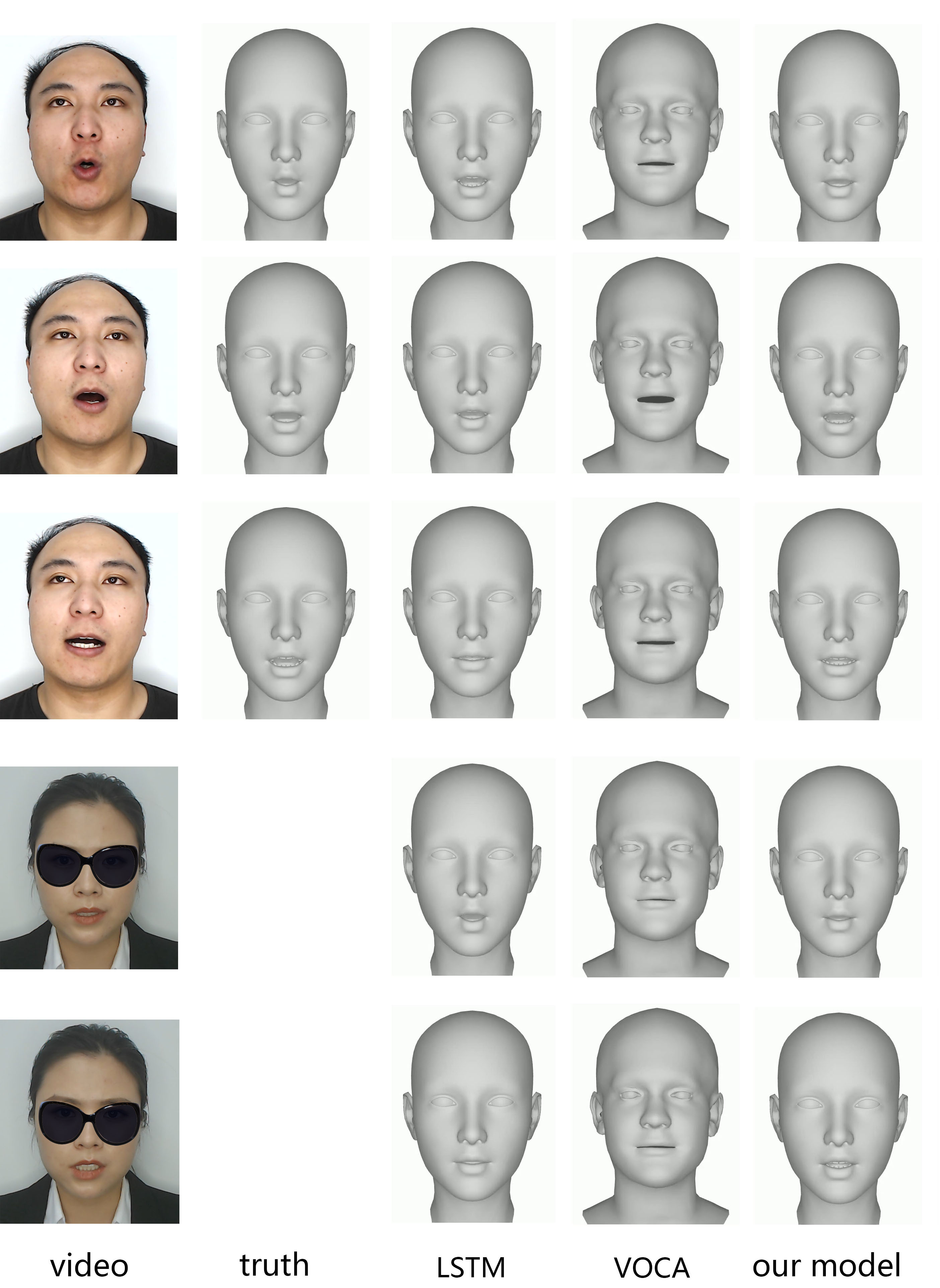}
\caption{Samples generated by different models. From left to right, the columns show the face appearance of the character, the ground truth of 3D face generated by Faceware, and results of LSTM, VOCA and our model respectively.}
\label{face-figure}       
\end{figure}

Figure~\ref{face-figure} shows some samples. The first column on the left shows the video frames of the character, and the second column is the ground truth of the 3D face which is reconstructed using Faceware. Facial animations generated by LSTM, VOCA and our model are shown in the third, fourth and the last columns respectively. The test speech of the first three rows comes from the test data set, which is the same male voice as the training data. The test speech of the last two rows is female voice and there is no ground truth of 3D face. It can be obviously seen that our model shows the best performance and generates more accurate mouth shapes than the other two methods.

\subsection{Quantitative Evaluation}\label{sec:res:quan}

Two metrics are used to measure the accuracy of generated facial animation and lip movements for different models: the positional error and the velocity error of the 2D facial landmarks. Two network structures with and without velocity loss are compared, and the results are shown in Table~\ref{error-table}.

\begin{table}
\centering
\caption{Error metrics of four models on the test data}
{\begin{tabular}{ccccc} \toprule
&\textbf{LSTM} & \textbf{LSTM+v\_loss} & \textbf{Conv+LSTM} & \textbf{Conv+LSTM+v\_loss} \\ \midrule
&\multicolumn{4}{c}{position error of the 2D facial landmarks (unit:pixel)} \\ \midrule
& 2.449 & 2.401 & \textbf{2.275} & 2.296 \\ \midrule
&\multicolumn{4}{c}{position error of the 2D mouth landmarks (unit:pixel)} \\ \midrule
& 4.920 & 4.812 & \textbf{4.514} & 4.592 \\ \midrule
&\multicolumn{4}{c}{velocity error of the 2D facial landmarks} \\ \midrule
& 4.022 & 3.593 & 3.719 & \textbf{3.397} \\ \midrule
&\multicolumn{4}{c}{velocity error of the 2D mouth landmarks} \\ \midrule
& 5.704 & 4.537 & 4.881 & \textbf{4.204} \\ \bottomrule
\end{tabular}}
\label{error-table}
\end{table}

It can be seen from the qualitative results that the combination of convolution layers with LSTM achieves higher action accuracy than the application of LSTM alone, both on the result of the landmark position error and that of the motion error. The comparison results of the velocity loss function is a bit complicated. For the LSTM network, the introduction of the velocity loss function has shown positive effects both on positional error and motion error. For the network with the combination of convolution layers and LSTM blocks, the velocity loss function reduces the motion error of the vertices, but it does not decrease the positional error of the landmarks. Actually, the positional error of the landmarks can only represent the difference between the facial shape output of the network and the ground truth, rather than the reality and naturalness of the facial animation. For example, the amplitude of the mouth movement is different when different people speak. The mapping between the speech and the face shape is not unique. Therefore, it is not appropriate to judge the contribution of the velocity loss function to the fidelity of the facial animations from the positional error of landmarks. On the other hand, as the velocity error of the landmarks is able to reflect the error of facial motion, we believe that the introduction of the velocity loss function is beneficial to the network we proposed.

\begin{figure}
\centering
\includegraphics[width=0.4\textwidth]{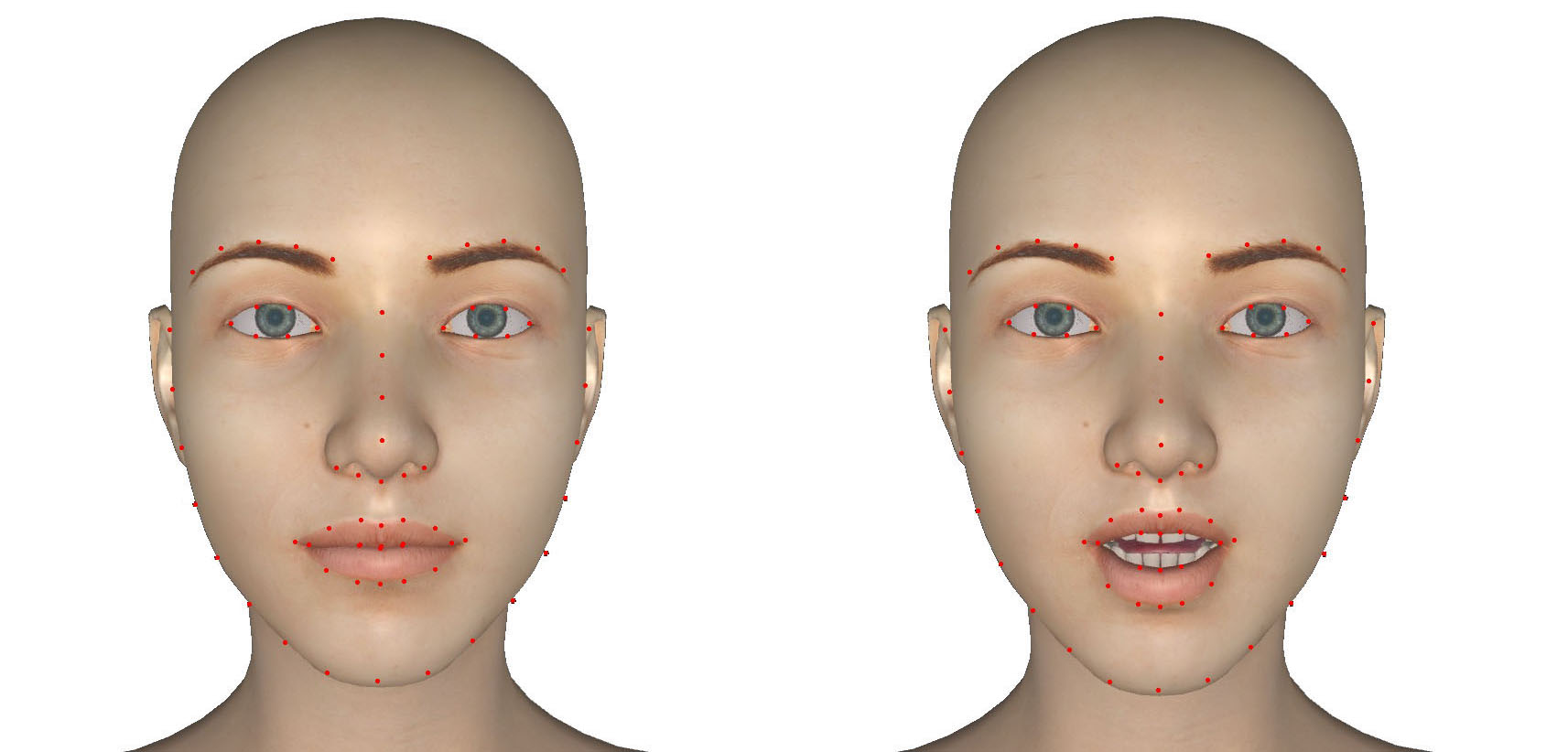}
\caption{Rendered facial animations and the detected 2D facial landmarks.}
\label{landmark-figure}       
\end{figure}

\begin{figure*}
\centering
\includegraphics[width=0.8\textwidth]{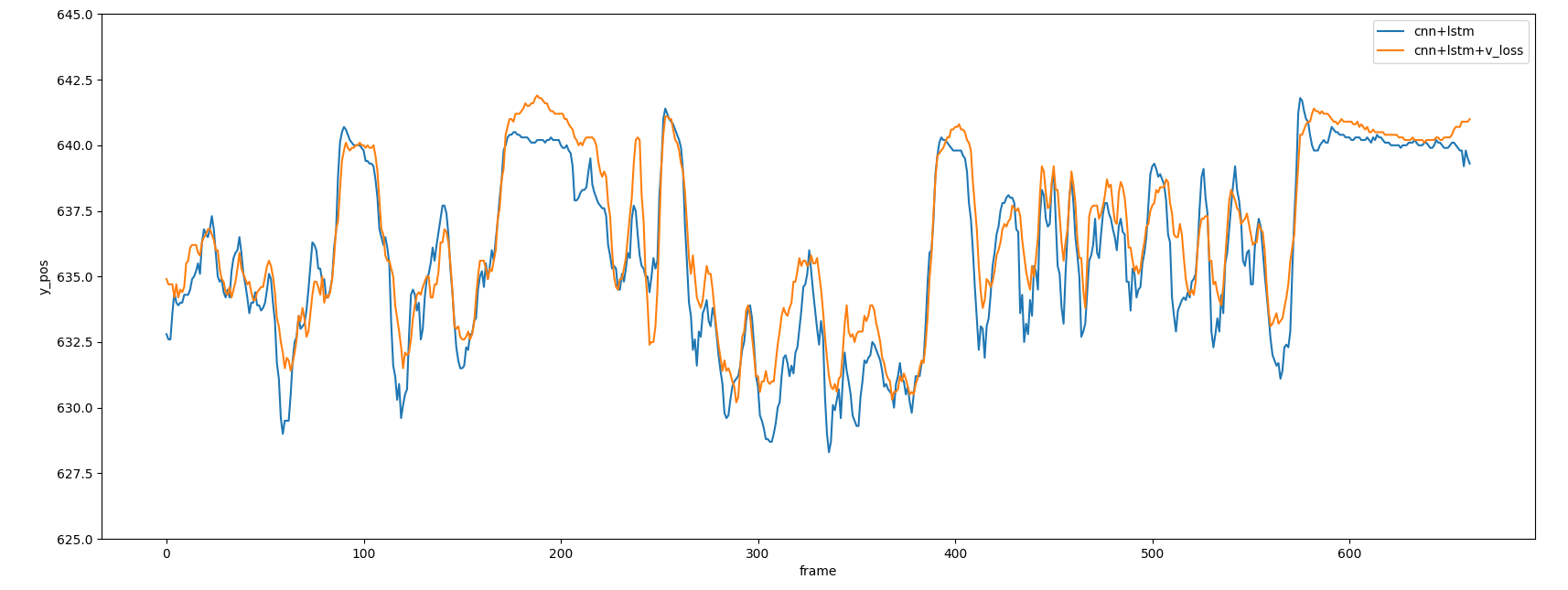}
\caption{Movement curves of the selected lip feature point generated by our models trained with and without the velocity loss term.}
\label{vloss1-figure}       
\end{figure*}

\begin{figure*}
\centering
\includegraphics[width=0.85\textwidth]{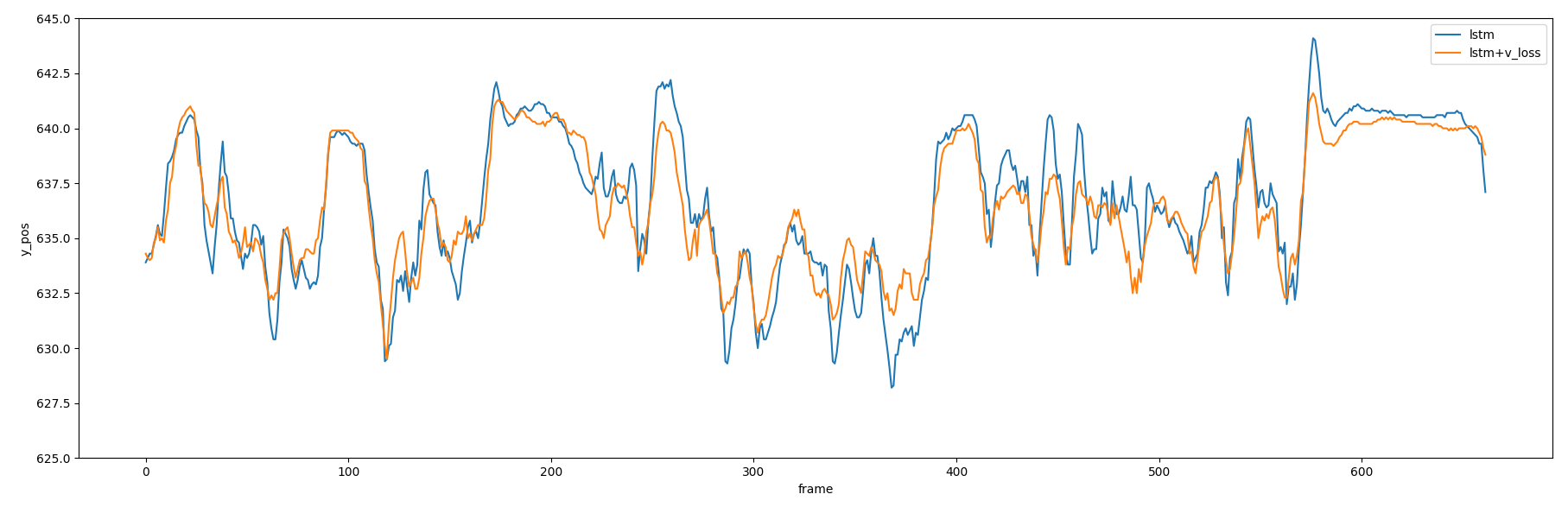}
\caption{Movement curves of the selected lip feature point generated by LSTM models trained with and without the velocity loss term.}
\label{vloss2-figure}       
\end{figure*}

We also conducted experiment to prove if the velocity loss we added is able to reduce lip jitter and make the animation transition between words much smoother. Similar to~\cite{Richard 2021}, we render 3D facial animations into images and detect 2D facial landmarks with a facial behavior analysis toolkit OpenFace~\cite{Baltrusaitis 2018} (see Figure~\ref{landmark-figure}). We extract the positions of a feature point located in the middle of the upper lip, and draw the movement curve of this feature point in the $y$ direction. Figure~\ref{vloss1-figure} and Figure~\ref{vloss2-figure} show the movement curves generated by models with and without the velocity loss term respectively. The base model adopted in Figure~\ref{vloss1-figure} uses our network architecture which combines convolutional layers with LSTM blocks, and that adopted in Figure~\ref{vloss2-figure} uses the LSTM network. It can be seen from the two figures that the introduced velocity loss term significantly reduces the lip jitter and smooths the lip movement for both networks. It is also noted that the velocity loss reduced the motion range of the mouth. Therefore, it is necessary to comprehensively consider the trade-offs to select an appropriate weight for the velocity loss term.

\section{Conclusion}\label{sec:con}

In this work, we created a new speech-animation dataset which contains the video and animation data of an adult speaking Mandarin, and proposed a novel network for the generation of facial animation from speech. We conducted both quantitative and qualitative evaluations to show that our network is able to generate more accurate and smooth facial animation, and is more robust to various audio sources. It can be further improved by incorporating more movement data of the upper face part, such as eyebrows and eyes, to make the facial animation of digital human more natural and realistic.






%

\end{document}